\magnification=\magstep 1
\noindent
{\bf Decoherence of Macroscopic Closed Systems within Newtonian Quantum
\break Gravity}
\bigskip
\bigskip

\noindent
Bernard S. Kay\footnote{$^*$}{e-mail bsk2@york.ac.uk}
\bigskip
\noindent
Department of Mathematics, University of York, York YO10 5DD, UK
\bigskip
\bigskip

\beginsection Abstract

A theory recently proposed by the author aims to explain decoherence and
the thermodynamical behaviour of closed systems within a conservative,
unitary, framework for quantum gravity by assuming that the operators
tied to the gravitational degrees of freedom are unobservable and
equating physical entropy with matter-gravity entanglement entropy. 
Here we obtain preliminary results on the extent of decoherence this
theory predicts. We treat first a static state which, if one were to
ignore quantum gravitational effects, would be a quantum superposition
of two spatially displaced states of a single classically well
describable ball of uniform mass density in empty space. Estimating the
quantum gravitational effects on this system within a simple Newtonian
approximation, we obtain formulae which predict e.g. that as long as the
mass of the ball is considerably larger than the Planck mass, such a
would-be-coherent static superposition will actually be decohered
whenever the separation of the centres of mass of the two ball-states
excedes a small fraction (which decreases as the mass of the ball
increases) of the ball radius.  We then obtain a formula for the quantum
gravitational correction to the would-be-pure density matrix of a 
non-relativistic many-body Schr\"odinger wave function and argue that this
formula predicts decoherence between configurations which differ (at
least) in the `relocation' of a cluster of particles of Planck mass.  We
estimate the entropy of some simple model closed systems, finding a
tendency for it to increase with `matter-clumping' suggestive of a link
with existing phenomenological discussions of cosmological entropy
increase.

\bigskip     
\bigskip     

\noindent     
In [1] a theory was proposed for the origin both of decoherence and of
thermodynamics in which quantum gravity plays a fundamental role. The
starting point is the following, conservative, set of assumptions:  Any
closed quantum gravitational system is described by a total Hilbert
space $H_{\hbox{total}}$ which takes the form 
$$H_{\hbox{total}}=H_{\hbox{matter}}\otimes H_{\hbox{gravity}}$$     
and any full description of the system is given, at all times, by a pure
density operator  $$\rho=|\Psi\rangle\langle\Psi|$$ on
$H_{\hbox{total}}$. An `initial' such density operator $\rho_0$ at one
`instant of time'  is assumed to evolve according to a
Schr\"odinger-picture unitary time evolution   
$$\rho(t)=U(t)\rho_0 U(t)^{-1}$$   
where $U(t)$ is a function on the non-negative real numbers taking its
values in the unitary operators on $H_{\hbox{total}}$. 

Decoherence then arises as a consequence of adding the (new) assumption
that the operators which correspond to physical observables take the
form $A\otimes I$ where $A$ acts on $H_{\hbox{matter}}$ and $I$ is the
identity operator on $H_{\hbox{gravity}}$; in words, that {\it the
operators tied to the gravitational degrees of freedom are
unobservable}.  In this way, the expectation value of any such
observable in a pure total state $\rho$  will be given by the formula
$\hbox{tr}(\rho_{\hbox{matter}}A)$ where $\rho_{\hbox{matter}}$ is the
partial trace of $\rho$ over $H_{\hbox{gravity}}$. Whenever the total
state $\rho$ cannot be written as a single tensor product,
$\rho_{\hbox{matter}}$ will be a mixed state with a non-zero entropy,
given by the formula
$$S=-\hbox{tr}(\rho_{\hbox{matter}}\ln\rho_{\hbox{matter}})\eqno{(1)}$$  
(i.e. the `matter-gravity entanglement entropy' of $\rho$). The theory
is completed by the declaration that $S$ be identified with the physical
entropy of thermodynamics.

We recall from [1] that the existing evidence for this theory is that it
offers a natural explanation as to how it can be that the laws of black
hole mechanics are equatable with the ordinary laws of thermodynamics,
and also as to why the previously calculated `matter' and `gravity'
entropy-like-quantities associated with a quantum black hole turn out to
be equal.  Further, this theory goes together with a natural resolution of
the `information loss puzzle' [2,3].  In this resolution [1], the lost
information is stored in the form of matter-gravity correlations.  Also,
on the additional assumption that the `initial state' $\rho_0$ of the
universe was unentangled, it would seem to offer the prospect [1] of being
able to derive, as a deterministic prediction within a theory of quantum
gravity consistent with our assumptions, the result that the entropy of
the universe must always increase, and thus of finally providing a
satisfactory microscopic explanation for the second law of
thermodynamics.  

In this letter, we obtain some preliminary estimates\footnote{$^1$}{We
shall use Planck units, $G=c=\hbar=k=1$, throughout the paper so, in
particular, masses are in units of the Planck mass $\approx 2\times
10^{-5}$ grams.  We shall also take the metric $\eta_{ab}$ on Minkowski
space to be $\hbox{diag}(-1,1,1,1)$.}, in the context of non-relativistic
quantum mechanics, for the magnitude of gravity-induced decoherence
entailed by our theory and then examine the reasonableness of the claim
that physical entropy is given by the formula (1).  First we consider a
very simple model for a `Schr\"odinger-cat-like'  superposition: Instead
of a cat in a superposition of alive and dead states, we take (in a
description which temporarily ignores gravity) a ball of radius $R$,
mass $M$, and uniform mass density $\mu=3M/(4\pi R^3)$ which is in a
static superposition of two states which are each well-describable as
classical states at rest in a given frame in flat spacetime, one
centred, say, at position ${\bf x_1}$ and the other at  ${\bf x_2}$ in
the given frame. Schematically:   
$$|\hbox{state}\rangle= c_1 |\hbox{ball centred at}\ {\bf x_1}\rangle +
c_2 |\hbox{ball centred at}\ {\bf x_2} \rangle$$ 
with $|c_1|^2+|c_2|^2=1$.   We shall confine our discussion to radii and
mass densities consistent, in a classical description which includes
gravity, with a very nearly flat spacetime. 

It seems reasonable that, in a fundamental quantum description which
takes into account quantum gravitational effects but ignores the surely
tiny `back-reaction' of the gravitational field on the ball, the state
which would correspond to our ball in a single position would be
described by a $|\Psi\rangle$ in $H_{\hbox{total}}$ which takes the form
$$|\Psi\rangle=|B\rangle\otimes |\gamma\rangle$$  where $|B\rangle$ is
an element of $H_{\hbox{matter}}$ which corresponds, as closely as
quantum theory will allow, to our classical ball in the case one ignores
the gravitational field, while $|\gamma\rangle$ in $H_{\hbox{gravity}}$
is the quantum state of the gravitational field in the presence of the
ball.  At this level of description one thus expects that the schematic
equation above should be replaced by 
$$|\Psi\rangle=c_1 |B_1\rangle\otimes |\gamma_1\rangle + c_2
|B_2\rangle\otimes |\gamma_2\rangle\eqno{(2)}$$ 
where $|B_1\rangle$ and $|B_2\rangle$ now represent our ball in each of
the two positions  in the superposition and  $|\gamma_1\rangle$ and
$|\gamma_2\rangle$ are the corresponding gravity  states. Clearly, if,
in some limit, $|\gamma_1\rangle$ and $|\gamma_2\rangle$ were to become 
orthogonal to one another in $H_{\hbox{gravity}}$, then
$\rho_{\hbox{matter}}$ would tend to 
$$\rho_{\hbox{matter}}=|c_1|^2 |B_1\rangle\langle B_1|+ |c_2|^2
|B_2\rangle\langle B_2|,$$ 
i.e. to a state of complete decoherence.  Thus, in order to get a simple
estimate for the extent of decoherence, it suffices to calculate the
inner product $\langle\gamma_1|\gamma_2\rangle$.  If this has modulus
very close to 1 then there is almost no decoherence.  If it has modulus
very close to zero, then decoherence is almost complete.

With our assumptions of weak gravitational fields and motionless ball
states, classically, the state of the gravitational field for our ball
centred at the origin will be well described by the Newtonian potential
$\phi$,  vanishing at infinity, and satisfying
$$\nabla^2\phi=4\pi\mu\eqno(3)$$ 
where $\mu$ denotes the mass-density of the ball centred at the
origin. We shall need the solution to this in the form of the
Fourier transform: 
$$\tilde\phi({\bf k})=-4\pi{\tilde\mu({\bf k})\over
k^2}=-{4\pi\over(2\pi)^{3/2}k^2}\int\! \mu({\bf x}) e^{-i{\bf k}.{\bf
x}}\,d^3{\bf x}=-{6M\over \sqrt{2\pi}R^3}\left ({\sin kR-kR\cos kR \over
k^5}\right ).\eqno{(4)}$$  
The Fourier transforms of the Newtonian potentials, $\phi_1$, $\phi_2$ of
our balls centred at ${\bf x_1}$, ${\bf x_2}$ are then given by 
$\tilde\phi_1({\bf k})= 
e^{i{\bf k}.{\bf x_1}}\tilde\phi({\bf k})$, $\tilde\phi_2({\bf k})= 
e^{i{\bf k}.{\bf x_2}}\tilde\phi({\bf k})$.
We shall now obtain a formula for $\langle\gamma_1|\gamma_2\rangle$
under the approximation that one quantizes the gravitational field, but
continues to regard $\mu_1$ and $\mu_2$ as fixed classical sources. We
shall first obtain the appropriate formula ((7) below)  for the
(unphysical) case of spin-zero gravity, where the full dynamical equation
for the Newtonian potential $\phi$ is the scalar wave equation with source 
$$(\partial^2/\partial t^2-\nabla^2)\phi=-4\pi\mu,\eqno{(5)}$$  
since it will be easier to explain the calculation for its physically
correct spin-two analogue after that has been done.  We shall find that
the two results have the same functional form (compare (7) and (12)
below)  but that the spin-two case has a `decoherence exponent' (see
after (7)) $D_2$, six times as large as that for spin-zero.

Turning to the spin-zero calculation, we shall proceed under the
`single-Fock-space assumption' that one may describe quantum states
entirely within the vacuum representation of the corresponding
source-free quantum field $\hat\phi(x)$ defined, in the usual way, on
the Fock space over the one-particle Hilbert space $h$ consisting of
square integrable functions on momentum space, by setting 
$$\hat\phi(t,{\bf x}) =(2\pi)^{-3/2}\!\int\! (1/\sqrt{2|k|})(a({\bf
k})e^{-i|k|t+i{\bf k}.{\bf x}} + a^+({\bf k})e^{i|k|t+i{\bf k}.{\bf
x}})\,d^3{\bf k}.$$  
Here $[a({\bf k}), a^+({\bf k'})]=\delta^3({\bf k}-{\bf k'})$ and
$a({\bf k})$ annihilates the Fock vacuum vector which we shall denote
below by $|\Omega\rangle$.  We shall explain below that while this
assumption is not strictly correct mathematically, it nevertheless leads
to the correct result.

With this assumption, we take our quantum description $|\gamma\rangle$
of the state of the gravitational field for a ball centred at the origin
to be the coherent state
$$|\gamma\rangle=
e^{-\langle\psi|\psi\rangle_h/2}e^{a^+(\psi)}|\Omega\rangle\eqno{(6)}$$
where the one-particle wave function $\psi({\bf k})$ is defined to be 
$|k|^{1\over 2}\tilde\phi({\bf k})/\sqrt 2$ and  $a^+(\psi)$ means $\int\!
a^+({\bf k})\psi({\bf k})\,d^3{\bf k}$. One justification for this is
that, as one may easily check, the expectation value 
$\langle\gamma|\hat\phi(x)\gamma\rangle$ of the source-free quantum
scalar field is then equal to $\phi(x)$ while the higher truncated
n-point functions [4] are the same as in the source-free vacuum.  (Another
justification may be had from the remark about the canonical formulation
below.)

Clearly, the two quantum states $|\gamma_1\rangle$, $|\gamma_2\rangle$ 
in our superposition will be given by replacing $\psi({\bf k})$ in (6)
by  $\psi_1({\bf k})=e^{i{\bf k}.{\bf x_1}}\psi({\bf k})$ and
$\psi_2({\bf k})=e^{i{\bf k}.{\bf x_2}}\psi({\bf k})$. Thus from (6)
$$\langle\gamma_1|\gamma_2\rangle=|\langle\gamma_1|\gamma_2\rangle |
=\exp{(-D_0)}\eqno{(7)}$$ 
where $D_0$, which we name the {\it spin-zero decoherence exponent}, is
given by
$$D_0=\|\psi_1-\psi_2\|_h^2/2 \eqno{(8)}$$
which can easily be rewritten as
$$D_0=\langle\psi|(1-e^{i{\bf k}.{\bf a}})\psi\rangle_h\eqno{(9)}$$
where ${\bf a}={\bf x_2-x_1}$ is the displacement between our two
ball-states, and which, by (4) is given by the explicit formula 
$$D_0=36M^2\!\int_0^\infty  {(\sin\kappa-\kappa\cos\kappa)^2\over\kappa^7}
\left ({\kappa\alpha-\sin\kappa\alpha \over
\kappa\alpha}\right)\,d\kappa\eqno{(10)}$$
where $\alpha=a/R$.

To have a clear mathematical perspective on this calculation, one needs
to note that $\psi({\bf k})$ is not square integrable; there's a
logarithmic infra-red divergence.  Thus $|\gamma\rangle$ doesn't really
make sense as a vector state in the source-free field vacuum sector (and
similarly for $\psi_1$, $\psi_2$ and $|\gamma_1\rangle$ and
$|\gamma_2\rangle$).  Instead, it should be more properly regarded as a
state in the algebraic (see e.g. [4]) sense -- obtained by composing the
source-free vacuum state with the automorphism determined by the mapping
$\hat\phi \mapsto \hat\phi-\phi I$ where $\hat\phi$ denotes the quantum
field, $I$ is the identity operator and $\phi$ solves (3).  Nevertheless
the convergence of the integral (10) indicates to us that the two states
$|\gamma_1\rangle$, $|\gamma_2\rangle$ do belong to the same sector and
have inner product correctly given by (7).  We conclude that it is this 
sector which should then be identified with  $H_{\hbox{gravity}}$ in
this spin-zero model.  However, we have also been reassured that, as we
anticipated, our `single-Fock-space assumption', while incorrect from a
mathematically strict point of view, does lead to the correct result. 

As a final remark about the spin-zero case, we notice that, from the point
of view of canonical quantization, the above automorphism corresponds to
the classical canonical map $\pi\mapsto\pi$, $\varphi\mapsto\varphi-\phi$,
and the significance of this map is that it sends the Hamiltonian
$H_0={1\over 2}\!\int (\pi^2 +(\nabla\phi)^2)\, d^3{\bf x}$ of the
source-free theory into the `quadratic part' (i.e.  after `completing the
square') of the Hamiltonian $H_\mu={1\over 2}\!\int (\pi^2
+(\nabla(\varphi-\phi))^2-(\nabla\phi)^2)\,d^3{\bf x}$ for the case of
non-vanishing source.  (The `constant' part, $-{1\over 2}\!\int
(\nabla\phi)^2\, d^3{\bf x}$, plays no role in our calculation, although
of course it is of interest in that it equals the expectation value in our
coherent quantum state of the energy of the gravitational field.) 

Of course, assuming the correctness of classical general relativity, the
physically correct formulation of our approximate quantum theory for the
gravitational field, i.e. {\it true Newtonian quantum gravity}, should
be based, not on the spin-zero equation (5), but rather on the,
spin-two, theory in which the gravitational field is described by a
linearized metric perturbation $h_{ab}$, $a,b = 0,1,2,3$, in Minkowski
space with a dynamics determined by demanding that $\eta_{ab} + h_{ab}$
satisfy the linearized Einstein equations  $G^{\hbox{linear}}_{ab}=8\pi
\mu\delta_{a0}\delta_{b0}$.  We shall now explain how to obtain the
appropriate notion of coherent states for this theory, taking the above
remark about the canonical interpretation of our scalar-gravity coherent
states as a useful clue.  In the case that $\mu$ vanishes, one may take
the source-free Hamiltonian for spin-two gravity to be that for a
$3\times 3$ matrix of free massless scalar fields in Minkowski space
described by canonical variables, $(\pi_{ij}$, $\varphi_{ij})$,
$i,j=1,2,3$, where for correct normalization $\varphi_{ij}$ is to be
identified with $h_{ij}/\sqrt 2$, subject to the constraints that they
be symmetric (i.e. $\pi_{ij}=\pi_{ji}$ and $\varphi_{ij}=\varphi_{ji}$),
transverse (i.e. $\nabla_i\pi_{ij}=0$ and $\nabla_i\varphi_{ij}=0$), and
traceless (i.e. $\pi_{ii}=0$ and $\varphi_{ii}=0$).  It is
straightforward to now show that the spin-two analogue of our canonical
map for the spin-zero case is the map $\pi_{ij}\mapsto \pi_{ij}$, 
$\varphi_{ij}\mapsto \varphi_{ij}-\sqrt 2\phi\,\delta_{ij}$, where,
again, $\phi$ solves (3).   Indeed, one may check that this maps both
the source-free Hamiltonian into (the quadratic part of) the correct
Hamiltonian with source $\mu$, and the source-free constraints into the
correct constraints for source $\mu$. With the appropriate analogue of
our above single-Fock-space assumption, this easily leads to the
appropriate description of our spin-two coherent state as the vector,
now in the big Fock space each of whose elements is a $3\times 3$ matrix
of elements of the usual scalar field Fock space:
$$|\gamma\rangle=
\exp(-3\langle\psi|\psi\rangle_h)
\exp(a_{ii}^+(\sqrt 2\psi))|{\bf\Omega}\rangle\eqno{(11)}$$
where $\psi$ is defined as after equation (6), $|{\bf
\Omega}\rangle$ represents the vacuum vector in our big Fock space, and
$a_{ij}^+$ is the usual creation operator on the $ij$th matrix element
of the latter.  In words, one can thus think of $|\gamma\rangle$ in this
spin-two case as a coherent state of `transverse but non-tracefree
gravitons'. Using (11) it is easy to see that the correct spin-two
analogue of equation (7) is then
$$\langle\gamma|\gamma\rangle=|\langle\gamma|\gamma\rangle|
=\exp(-D_2)\quad\hbox{where}\quad D_2=6D_0\eqno{(12)}$$ 
thus establishing the result mentioned at the outset.

We now obtain some first predictions from (12).  In the case $a\ll
R$, one finds from (10) that 
$$D_2=6D_0=9M^2\left ({a^2\over R^2} + O({a^4\over R^4} \ln {a\over R})
\right )\eqno{(13)}$$  
thus predicting, for example that in the case of a ball with the density
of water and radius $0.1$ cm, $D_2\approx 4\times 10^7 A^2$ 
where $A$ now denotes the separation of the centres of mass measured in
centimetres.  Thus in this case one might say there is a {\it
decoherence length} of around $1.6\times 10^{-4}$ cm and we would need
the centres to be closer together than $10^{-5}$ centimetres or so in
order to be able to ignore gravitational effects in making quantum
predictions involving interference.   In the
case\footnote{$^2$}{Concerning the intermediate regime, from an analysis
of (10) using computer mathematics packages it appears, satisfactorily,
that, for fixed $M$ and $R$, $D_0$, and hence also $D_2$, increases
monotonically with $a$.} $a\gg R$, one finds from (10) that   
$$D_2=6D_0=24M^2\left(\ln(a/R) + O(1)\right )\eqno{(14)}$$   
thus predicting, for example, that for a ball with the density of water
and radius $7\times 10^{-3}$ cm,  the extent of decoherence will
increase from a small but noticeable to a large amount as the separation
of the centres of mass increases from, say $0.1$ cm to 10 metres.
Reassuringly, one obtains no significant violation of the superposition
principle for balls with masses and radii typical of elementary
particles (e.g. neutrons).   

It is worth pausing to compare the above results with the analogous
results one would obtain for superpositions of ball-states with electric
charge $Q$ on the (we presume false) assumption that the electromagnetic
field were unobservable.  Now one needs of course the spin-one analogue
of the above notion of coherent states.  Replacing $M^2$ above (measured
in units of Planck mass squared) by $Q^2$ (measured in units such that
the square of the charge of the electron $e^2$ is equal to the fine
structure constant $\approx 1/137$) and with an analysis similar to that
given above for the case of spin-two,  one finds that the canonical map
(now on the canonical variables $\pi_i$, equal to minus the electric
field strength, and $\varphi_i$) which maps the Hamiltonian and
constraints ($\nabla_i\pi_i=0$, $\nabla_i\varphi_i$=0) for the charge
free theory into those for the theory with charge is now
$\pi_i\mapsto\pi_i-\nabla_i\phi$, $\varphi_i\mapsto\varphi_i$.   From
this, one easily sees that the appropriate spin-one notion of coherent
state (which one finds may be regarded as consisting of `longitudinal
photons') leads to an analogue of equation (7) which is {\it identical
with} (7); i.e. we find a spin-one decoherence exponent $D_1=D_0$
$(={1\over 6}D_2)$. Thus one would predict e.g. that a macroscopic ball
of radius $0.1$ cm and with a uniformly distributed charge
totalling 1 e.s.u. would  have a `decoherence length' as little as
$4.5\times 10^{-10}$ cm.  Also a superposition of two proton states,
modelled by a uniformly charged ball with the usual proton radius $R_p$
and total charge $e$ would have a decoherence exponent $D_1 \approx
{4\over 137}\ln(a/R_p)$ for $a\gg R_p$. Thus for the proton (and
similarly for other charged elementary particles) one would predict a
possibly noticeable amount of `decoherence' at almost any separation and
large amounts of `decoherence' at distances approaching say a centimetre
or more.  However, we would rather call this electromagnetic analogue
`pseudo-decoherence' because, unlike the gravitational field on our
theory, we take the electromagnetic field to be observable.  But anyway,
returning to the case of macroscopic balls, it is interesting to note
that it should not be too difficult to have macroscopic balls which are
sufficiently electrically neutral for the gravitational decoherence our
formulae predict for their centre of mass to quite easily exceed in
magnitude the corresponding electromagnetic pseudo-decoherence.

We now discuss the general conclusions which may be inferred at this
stage and point out some relationships with other work. The general
conclusion strongly suggested by our results is that our theory
predicts, in the non-relativistic approximation adopted, that static
quantum superpositions of macroscopically different configurations
cannot exist as coherent superpositions and will, instead, spontaneously
and instantaneously decohere at the moment of attempted manufacture. 
Here, a rough quantitative interpretation of the phrase `macroscopically
different configurations', suggested by an examination of both (13) and
(14), is that:  

\medskip 
\noindent 
{\it the second configuration differs from the first (at least) in that
a lump of Planck mass or more has been `relocated' to a disjoint region
of space.}\qquad\qquad\qquad\qquad\qquad\qquad\qquad (15) 

\medskip
\noindent 
(Further support for (15) can be had on examination of (16), (17) below,
interpreting `lump' to mean `cluster of particles'.)

We remark that the instantaneous nature of the decoherence is presumably
an artefact of our use of a non-relativistic approximation.  Instead, it
seems reasonable to expect that in a more accurate, relativistic, analysis
of examples such as those above, one would find that it was possible to
prepare a coherent superposition but that this would decay with a time
scale set by the time for light to travel across our decoherence length.

The idea that gravity may play a fundamental role in decoherence has 
previously been suggested by a number of authors either in the context
of a linear, but non-unitary, framework [5,6] for the underlying quantum
mechanics or in the context of a possible gravity-related modification
of quantum mechanics itself [7,8,9,10] often implicitly or explicitly
involving the abandonment of linearity.   Especially the work of
Penrose, see [10] and references therein, appears to bear a particularly
close and interesting relationship to the present work as we now
discuss. [10] makes no attempt to specify the underlying theory,  but
instead draws the inference that `spontaneous state-vector reduction' of
macroscopic superpositions must take place on the basis of a fundamental
ill-definedness in the notion of `stationary state' when one attempts to
adopt the viewpoint of classical general relativity for the description
of a quantum superposition. Starting from a starting point similar to
our equation (2) for a general `lump' of mass density $\mu$  (cf. the
first equation in Section 4 of [10]), Penrose arrives, by a route which
is very different from that we have taken here, at a quantity called
$\Delta$ in [10], which in our notation can be written 
$$\Delta=\langle
(\phi_1-\phi_2)|-\nabla^2(\phi_1-\phi_2)\rangle$$ 
where the inner product denotes the ordinary $L^2$ inner product in
position space, and which, it is proposed, corresponds to a `spontaneous
state-vector reduction' time of the order of $1/\Delta$. This is to be
compared and contrasted with our quantity $D_2$ (12) (now, say, for the
same general lump) which, by [8], can be written, with the same notion
of inner product, as
$$D_2={3\over 2}\langle (\phi_1-\phi_2)|
\sqrt{-\nabla^2}(\phi_1-\phi_2)\rangle$$
which differs from $\Delta$ only in the square-root sign and in the
factor of $3/2$, and which, on our theory, corresponds to a decoherence
length (or, with the expectations mentioned above about the results in a
proper relativistic treatment, decoherence time) defined to be the
distance at which $D_2$ attains, say, the value 1. To summarize, while
there are interesting differences between these two predicted measures,
the present theory provides results on decoherence which are
remarkably similar to those anticipated for `spontaneous state vector
reduction' in [10] of a yet to be discovered theory.

One may generalize the above analysis of our two ball-state model to the
case of a classical ball whose centre of mass is in a general
superposition described by a general Schr\"odinger wave function.
Indeed, one can generalize further to the case of a non-relativistic
{\it collection} of many balls, to be interpreted below as atomic
nuclei, say for simplicity all identical.  This would be described, if
one ignores gravitational effects, say by a (suitably antisymmetric or
symmetric) wave function $\Psi$, of the centre of mass coordinates ${\bf
x_1}, \dots , {\bf x_N}$ of our N balls tensor producted with the
N-fold tensor product of N copies of the wave function $|M\rangle$ for a
single ball centred on the origin. To take into account gravitational
effects in a non-relativistic approximation, similar reasoning to that
we followed in the case of the two ball-state model then leads us to
replace  this $\Psi$ by the (entangled) total matter-gravity state
vector
$$\Psi({\bf x_1},\dots, {\bf x_N})|\gamma({\bf x_1},\dots, {\bf x_N};{\bf
k})\rangle$$
where we have now suppressed the trivial dependence on $|M\rangle$ and
represent an element of the total matter-gravity Hilbert space as a
function from our N-centre configuration space taking its values in
$H_{\hbox{gravity}}$ and where $|\gamma({\bf x_1},\dots, {\bf x_N};{\bf
k})\rangle$ denotes the Newtonian gravitational coherent state (11) with
$\psi({\bf k})$ replaced by a $\psi({\bf k})$ defined to equal
$\sqrt{k/2}$ times the Fourier transform of the sum of the classical
Newtonian potentials due to our classical balls centred at ${\bf
x_1},\dots, {\bf x_N}$.  It is straightforward to then see that the
density matrix $\rho_{\hbox{matter}}$ obtained by tracing over
$H_{\hbox{gravity}}$ the projector onto the above matter-gravity wave
function  is given by
$$\rho_{\hbox{matter}}({\bf x_1},\dots,{\bf x_N};{\bf x'_1},\dots,{\bf
x'_N}) =\qquad\qquad\qquad\qquad\qquad\qquad\qquad\qquad\qquad$$
$$\qquad\qquad\qquad\qquad\Psi({\bf x_1},\dots,{\bf
x_N})\Psi^*({\bf x'_1}, \dots,{\bf x'_N})\exp(-D_2({\bf x_1},\dots,{\bf
x_N};{\bf x'_1},\dots,{\bf x'_N})) \eqno{(16)}$$ 
where (by (8), and (12)) $D_2$ is now equal to
$3\|\psi-\psi'\|_h^2$  with $\psi'({\bf k})$ defined in the same way
that $\psi({\bf k})$ is defined, but with ${\bf x_1},\dots, {\bf x_N}$
replaced by ${\bf x'_1},\dots, {\bf x'_N}$.   Explicitly,  using
the same asymptotic estimate used to obtain (14),
$\exp(-D_2)$ may be estimated as 
$$\exp(-D_2({\bf x_1},\dots , {\bf x_N};{\bf x'_1},\dots , {\bf
x'_N}))\approx \prod_{i=1}^N\prod_{j=1}^N\left ({|{\bf x'_i}-{\bf
x_j}||{\bf x_i}-{\bf x'_j}|\over |{\bf x_i}-{\bf x_j}||{\bf x'_i}-{\bf
x'_j}|}\right )^{-12M^2}\eqno{(17)}$$ 
where one replaces the terms in the denominator by $R$ when $i=j$ (and
we ignore the tiny region of configuration space $\times$ itself 
where any other of the terms becomes smaller than, say, a thousand $R$).

It now becomes possible to study the decoherence and thermodynamic
properties predicted by our theory for a non-relativistic model closed
system of ordinary matter  described, in the usual way, by a
Schr\"odinger wave function $\Psi(t; {\bf x_1}, \dots, {\bf x_N})$ for a
collection of nuclei and electrons evolving according to electrostatic
(Coulomb) and gravitational (Newtonian) potentials.  On our theory, this
needs to be interpreted, at each instant of time, via the density matrix
(16) or rather its generalization to several species of particles (but
one would expect the dominant decoherence effects to be given by the
nuclei because of their much larger mass) and will have a time-varying
entropy given by (1).

In addition to the defect that this model ignores photons etc., we have
no reason to expect such a model closed system to exhibit realistic
thermodynamic behaviour since: (A) Our model will need to be restricted
to be sufficiently small not to suffer gravitational collapse to a state
where our assumptions of slow velocities and weak gravitational fields
no longer apply and presumably [11] it is just such situations of
collapse which are the main generators of entropy in the actual
universe. (B) There is no reason to expect it to be legitimate to regard
an actual, duly restricted in size, system of ordinary matter as a
closed system with its own matter wave function because it would, from
its past and ongoing interactions, be expected to be considerably
entangled with much larger regions of the universe.  Nevertheless, with
due caution as to the interpretation of the results, it is clearly of
interest to study the entropy (1) of density matrices (16) for some
simple model many-body closed systems and we have initiated such a
study. As a useful general guide, we expect that: 

\medskip 
\noindent 
{\it the entropy (1) of the density matrix (16) is crudely estimatable
as the logarithm of the `weighted' (i.e. with $|\Psi({\bf x_1}\dots {\bf
x_N})|^2$) maximum number of cells into which configuration space can be
divided with the property that when $\exp(-D_2)$ is evaluated between
typical configurations in any pair of non-neighbouring cells it is
negligibly small.}\qquad\qquad\qquad (18) 
\medskip

Using this principle together with a variety of further reasoning which
we shall outline as required, we find the following: 

\smallskip
\noindent 
1) For a wave function consisting of $N$ non-interacting bosons of mass
$M$ treated as uniform density balls of radius $R$, in an N-fold tensor
product of a single plane wave state in a cubical box of side $L$ with
periodic boundary conditions, we have calculated the entropy (1), on the
assumption that $NM^2 \ll 1$, by approximating $\exp(-D_2)$ in (16) by 
$1-D_2$ and explicity diagonalizing the resulting density matrix and
estimating the resulting sum as an integral, and find a leading
behaviour for $S$ equal to a slowly varying function of $L$ and $R$
(explicitly  $-c_1\ln(L/R)\ln(c_2 NM^2(R/L)^{3/2})$ where $c_1$ and
$c_2$ are constants of order one) times $NM^2$.  Thus, with particle
masses of the order of typical nuclear masses and e.g. a total mass of,
say,  $10^{6}$ grams, one finds a tiny total entropy of the order of 1
or less for any reasonably sized box.  One expects such a result to
apply to any such fully delocalized `gas-like state' of this mass or
less since one expects `likely' (i.e. weighted with $|\Psi|^2$) mass
fluctuations to be of the order of $N^{1\over 2}M$ which will, for such
masses, be very much less than the Planck mass and hence, by combining
the rules of thumb (15) and (18), produce only a tiny bit of entropy. 

\smallskip 
\noindent 
2) On the other hand, if one imagines a state in which the same total
quantity of matter is condensed into the form of a delocalized gas of
$n$ `clumps' each of Planck mass or more, then, applying the rule of
thumb (15) (and ignoring any `internal' entropy -- see paragraph 3
below) one expects the entropy to equal $n$ times a slowly varying
function of order a `small number' of box size and clump size. In fact,
for spherical clumps, the density matrix (16) will,  by the estimate
(13), resemble that of a thermal state of a gas of $n$ clumps at a
temperature $T=9M/(2R^2)$ where $M$ is now the clump mass and $R$ the
clump radius. Thus, with Planck-mass-sized clumps and a reasonably sized
box, our $10^6$ gram example would now acquire an entropy of, say,
$10^{12}$ to within an order of magnitude or so, which is very much
larger than that for the uncondensed gas, although of course still much
smaller than the entropy which would be typical of an actual everyday
thermodynamic (open!) system of this total mass, which (forgetting
photons) would be of the order of the number of nuclei or around
$10^{28}$. 

\smallskip 
\noindent 
3) If, instead, one imagines the same total quantity of matter condensed
into a single large (but small compared to box size) clump, then the
entropy due to the (we shall assume) delocalized centre-of-mass motion
of the clump would be expected, by combining (15) and (18), to be
approximately given by $S=c_1\ln(c_2ML/R)$ where $c_1$ and $c_2$ are
constants of order 1, $M$ is the clump mass, and $R$ now represents a
suitable notion of `clump diameter'.  For reasonable box sizes, this
value can never get very big. However, for a sufficiently large clump,
one expects there to be a much more important, `internal', contribution
to its entropy due to the delocalization of the part of the wave
function describing the relative motion of its constituent nuclei. To
see this, consider a (we shall pretend, simple cubic, monatomic) cubic
crystal of side $\ell$ in its ground state. Then, letting $\Delta\ell$
denote the uncertainty in $\ell$ due to zero-point lattice fluctuations,
and $\rho$ the crystal density, clearly a sufficient condition for `mass
relocations' (cf. (15)) of Planck mass or more is that the inequality
$\rho\ell^2\Delta\ell > 1$ holds.  Estimating $\Delta\ell$ to be
$({\ell/4\pi ms})^{1\over 2}$ where $m$ is the nucleon mass and $s$ is
the longitudinal speed of sound, one finds that this inequality will be
satisfied provided $\ell$ exceeds the critical value $\ell_0=(4\pi
ms/\rho^2)^{1\over 5}$. This depends a bit on the atom(s) the crystal is
made out of, but turns out typically to be around 1 cm. Suppose now our
single large clump were such a crystal of side $R$, then a naive
argument based on  thinking of it as consisting of lots of subcrystals
of side $\ell_0$, each of which can decohere in one of two ways,
suggests that the `vibrational contribution to its
entropy'\footnote{$^3$}{There will also be a contribution to the entropy
from delocalization of the piece of the wave-function describing
rotational motion of the crystal about its centre of mass but one easily
argues this will be small.} will be around $(R/\ell_0)^3\ln 2$.   Thus
we estimate the entropy of our $10^6$ gram example to be $10^5$ to within
an order of magnitude. In conclusion, also a single large clump will
have a much larger entropy than the gas-like state of paragraph 1 of
similar mass -- albeit not so large as that of the gas of many
Planck-mass-sized clumps of paragraph 2.  But note that the `internal
entropy' might well be much larger for a single large clump with much
more internal energy than a crystal in its ground state.  Finally, let
us remark that in all three examples, we found that our
estimate for the entropy scales roughly linearly with total mass.

\smallskip
 
In all these models, the values we obtain for the entropy are far
smaller than typical values which occur in ordinary thermodynamics. 
However, we gave reasons above as to why this is not to be regarded as
unexpected for the small non-relativistic models of closed systems
considered here.  Instead, we regard it as encouraging that we have
found evidence, in the context of these models, that entropy as we
define it has a tendency to increase with `matter clumping'.  This is
suggestive of a link with existing discussions of cosmological entropy
production based on a phenomenological notion of entropy -- see
especially the account due to Penrose in [11] -- according to which it
is just such matter clumping which plays the key role.  

Taken together with the satisfactory situation discussed above for
decoherence, we feel that these results strengthen the evidence for the
correctness of the theory proposed in [1].

\bigskip
\noindent
I am grateful to Roger Penrose for a helpful discussion.  I thank Simon
Eveson for assistance with computer integration. I thank Wanda Andreoni
and Alessandro Curioni for valuable comments on a draft of this work.

\beginsection References

\item{[1]} Kay B S 1998 {\it Entropy defined, entropy increase and
decoherence understood, and some black-hole puzzles solved} 
hep-th/9802172 (to appear)

\item{[2]} Hawking S W 1976 {\it Phys. Rev.} {\bf D14} 2460

\item{[3]} Preskill J 1992 {\it Do black holes destroy information?}
hep-th/9209058

\item{[4]} Haag R 1992 {\it Local Quantum Physics} (Berlin: Springer)

\item{[5]} Hawking S W 1982 {\it Commun. Math. Phys.} {\bf 87} 395

\item{[6]} Ellis J, Mavromatos N E and Nanopoulos D V 1997 {\it Vacuum
fluctuations and decoherence in mesoscopic and microscopic systems}
quant-ph/9706051

\item{[7]} K\'arolyh\'azy F 1966 {\it Nuovo Cimento} {\bf A42} 390 

\item{[8]} Di\'osi L 1989 {\it Phys. Rev.} {\bf A40} 1165 

\item{[9]} Ghirardi G C, Grassi R and Rimini A 1990 {\it Phys. Rev.}
{\bf A42} 1057

\item{[10]} Penrose R 1996 {\it General Relativity and Gravitation} {\bf 28}
581

\item{[11]} Penrose R 1979 {\it Singularities and time asymmetry} In
Hawking S W and Israel W (eds) {\it General Relativity, An Einstein
Centenary Survey} (Cambridge: Cambridge University Press)

\bye